\documentstyle[aps,prl,multicol,epsf]{revtex}

\renewcommand{\narrowtext}{\begin{multicols}{2} \global\columnwidth20.5pc}
\renewcommand{\widetext}{\end{multicols} \global\columnwidth42.5pc}

\def\eqa{\begin{eqnarray}}
\def\eea{\end{eqnarray}}
\newcommand{\be}{\begin{equation}}
\newcommand{\ee}{\end{equation}}

\begin{document}
\draft
\title{Staggered currents in the mixed state}
\author{Qiang-Hua Wang$^{1,2}$, Jung Hoon Han$^1$, and Dung-Hai Lee$^1$}
\address{$^{1}$Department of Physics,University of California
at Berkeley, Berkeley, CA 94720, USA}
\address{$^{2}$Physics Department and National Laboratory of Solid State
Microstructures, Institute for\\
Solid State Physics, Nanjing University, Nanjing 210093, China}
\maketitle

\begin{abstract}
The current pattern in the mixed state of high-$T_{c}$ superconductors is
studied in the U(1) mean field theory of the t-J model. Our findings are the
following. 1) In the absence of antiferromagnetism a robust staggered
current pattern exists in the core of vortices if the doping is not too
high. 2) At a fixed doping and with increasing magnetic field, the size of
the staggered current core expands, and eventually percolates.
3) The polarity of the staggered current is pinned by the
direction of the magnetic field.
4) Vortex cores locally modify the hole density - in a staggered
(non-staggered) core, the excess charge is slightly negative (positive). 5)
Gutzwiller projection does not wash out the staggered current. Finally we
present two experimental predictions concerning neutron scattering and STM
spectra that capture the signature of the staggered current induced by the
vortices.
\end{abstract}


\narrowtext Recently, there has been a renewal of interest in the so-called
staggered flux phase of the t-J model\cite
{kotliar,affleck,patrick,clmn,wang,sudip,nayak} where a circulating current
produces a staggered orbital magnetic moment. According to the mean-field
theory,\cite{kotliar,affleck} the staggered flux phase, although never
stable, is a close competitor of d-wave superconductivity when
antiferromagnetism is absent. This situation can be aptly described by a
Ginzburg-Landau (GL) theory with two competing order parameters.\cite{clmn}
A serious concern is whether the staggered flux phase actually exists in the
high-$T_{c}$ phase diagram.\cite{clmn,wang,sudip,nayak,mook,keimer}

Even if the staggered flux phase does not exist, GL theory would predict
that in the core of a vortex, where the superconducting order is suppressed,
the staggered flux order has a chance to appear. This is first pointed out
forcefully in a recent paper by Lee and Wen\cite{leewen} from the viewpoint
of their SU(2) mean field theory of the t-J model.
Recently we have shown that the staggered current
core also exists in the U(1) mean-field solution of the same
model.\cite{han} Moreover we
demonstrated that such current pattern survives the Gutzwiller
projection which removes any double occupation in the mean field results.

The purpose of this paper is to address several important issues concerning
the staggered current in the mixed state of high-$T_{c}$ superconductors.
One of our main results is the prediction for two experiments which are
relevant to the existence of staggered current in vortex cores. Our
conclusion is based on the U(1) mean-field solution and its Gutzwiller
projection.

Our starting point is the t-J model with Coulomb interaction,
\begin{eqnarray}
H &=&-t\sum_{\langle ij\rangle \sigma }(C_{i\sigma }^{\dagger }C_{j\sigma
}e^{-iA_{ij}}+{\rm h.c.})+J\sum_{\langle ij\rangle }({\bf S}_{i}\cdot {\bf S}%
_{j}-\frac{n_{i}n_{j}}{4})  \nonumber \\
&&+\frac{V}{2}\sum_{i\neq j}\frac{1}{r_{ij}}n_{i}n_{j}-\mu \sum_{i}n_{i}.
\label{ham}
\end{eqnarray}
In the above $A_{ij}=(2\pi e/hc)\int_{{\bf R}_{i}}^{{\bf R}_{j}}{\bf A}\cdot
d{\bf l}$ is the link phase produced by the physical vector potential ${\bf A%
}$. All other notations are standard. In the rest of the paper we use the
parameter set $t=V=3J$.

In the U(1) slave-boson approach the following replacements are made in Eq.~(%
\ref{ham}): 1) $C_{i{\sigma }}^{\dagger }C_{j{\sigma }}\rightarrow
f_{i\sigma }^{\dagger }f_{j\sigma }b_{j}^{\dagger }b_{i}$, 2) ${\bf S}%
_{i}\rightarrow (1/2)f_{i\alpha }^{\dagger }{\bf \sigma }_{\alpha \beta
}f_{j\beta }$, , 3) $n_{i}\rightarrow 1-b_{i}^{\dagger }b_{i}$, and 4) the
occupation constraint $\rightarrow b_{i}^{\dagger }b_{i}+\sum_{{\sigma }}
f_{i{\sigma }}^{\dagger }f_{i{\sigma }}=1$. In the above $b_{i}$ is the
boson and $f_{i{\sigma }}$ is the fermion annihilation operator
respectively. Our calculation is performed at zero temperature where the
bosons are assumed to have condensed. The spin-exchange term is replaced by
the following mean-field decoupling
\begin{eqnarray}
{\bf S}_{i}\cdot {\bf S}_{j} &\rightarrow &-\frac{3}{8}\sum_{\sigma
}[K_{ij}^{\ast }f_{i\sigma }^{\dagger }f_{j\sigma }+{\rm h.c.}]  \nonumber \\
&&-\frac{3}{8}[\Delta _{ij}^{\ast }(f_{i\downarrow }f_{j\uparrow
}-f_{i\uparrow }f_{j\downarrow })+{\rm h.c.}]  \nonumber \\
&&+\frac{3}{8}(|K_{ij}|^{2}+|\Delta _{ij}|^{2}),
\end{eqnarray}
where $K_{ij}$ and $\Delta _{ij}$, the mean-field hopping and pairing
amplitudes, are determined self-consistently, together with the condensate
boson order parameter $\langle b_i\rangle$ and the Lagrange multiplier that
enforces the occupancy constraint. Due to the internal U(1) symmetry we can
always choose the gauge so that $\langle b_{i}\rangle $ is real and
positive. It is important to stress that the above decoupling excludes the
spin magnetic moment $\langle {\bf S}_{i}\rangle $ as an order parameter,
and hence will not be
able to describe low doping regime in which commensurate/incommensurate spin
magnetic order exists. Aside from the above restriction, no other constraint
is placed on the mean-field parameters. The actual calculation is performed
numerically on a $N_{x}\times N_{y}$ lattice under twisted boundary
condition. Much of our treatment of the vortex lattice is the same as that
in Ref.\cite{macdonald}. The average magnetic flux density is $f=\frac{1}{%
N_{x}N_{y}}$ flux quanta, in units of $hc/e$, per plaquette.

The flux density $f$ causes the pairing amplitude $\Delta _{ij}$ to wind by $%
4\pi $ around each $N_{x}\times N_{y}$ unit cell. The question concerning
whether it is energetically more favorable to nucleate two $hc/2e$ vortices
or a single $hc/e$ vortex in the unit cell has been raised in the
literature. \cite{leewen,fluxquantum} {\it Our result shows unambiguously
that it is the $hc/2e$ vortex that is favored.} More specifically we have
checked that even when the initial condition corresponds to a single $hc/e$
vortex, {\it i.e.} two $hc/2e$ vortices on top of each other, the final
self-consistent solution always exhibits two separated $hc/2e$ vortices. We
emphasize, however, that such conclusion will be subject to change if the
strength of the Coulomb potential is modified. The reason for that is
because the vortex core is charged!

In Figs.1(a) and (b) we show two dimensional map of the hole density in the
unit cell (with two vortices) at doping levels $x=10\%$(a), and $15\%$(b).
The blue region marks lower and the red region marks higher hole density
respectively. Thus the vortex core is negatively charged in (a) and
positively charged for (b). The charge difference between the $x=10\%$
and the $x=15\%$ vortices turns out to be related to  the presence/absence
of staggered currents within the vortex core as we will discuss below.

\begin{figure}
\epsfxsize=8cm\epsfysize=6.5cm\epsfbox{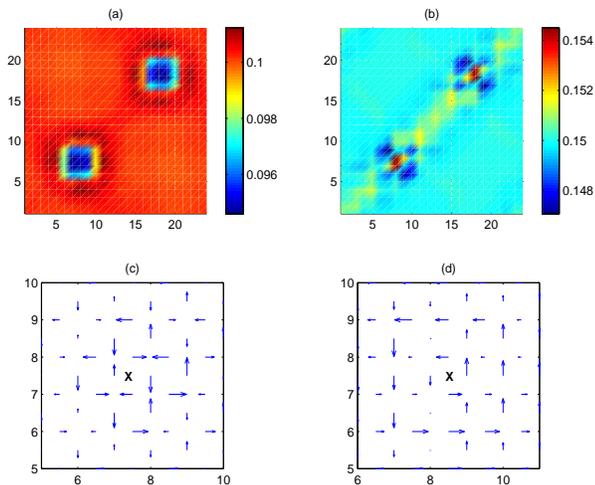}
\caption{Two vortices in a $24\times 24$ lattice. (a) Hole distribution with
$x=10\%$; (b) Hole distribution at $x=15\%$; (c) Bond current pattern
corresponding to (a); (d) Bond current pattern corresponding to (b).
The crosses in (c) and (d) highlight the vortex cores. }
\end{figure}

In Fig.1(c) we present the current pattern near the core of the lower-left
vortex in Fig. 1(a). The variable-length arrows represent the direction and
magnitude of the bond current given by
\begin{equation}
\langle J_{ij}\rangle =-i\langle b_{i}\rangle \langle b_{j}\rangle \langle
f_{i{\sigma }}^{\dagger }f_{j{\sigma }}{\rm e}^{-iA_{ij}}\rangle +{\rm c.c.}
\end{equation}
(Recall that in our gauge the boson condensate amplitude $\langle
b_{i}\rangle $ is real and positive.) The pattern clearly indicates the
existence of a staggered current core. In Fig.1(d) we present the current
pattern near the core of the lower-left vortex in Fig.1(b), where the
staggered current has disappeared.

\begin{figure}
\epsfxsize=8cm\epsfysize=7cm\epsfbox{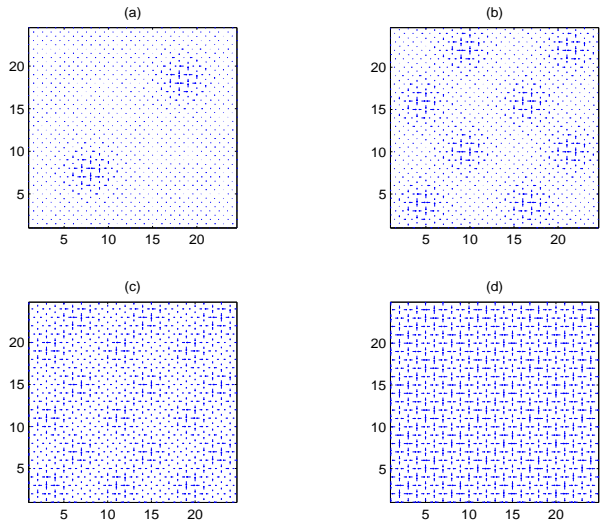}
\caption{Bond current patterns at (a) $f=1/576$, (b) $1/144$, (c) $1/64$,
and (d) $1/36$.}
\end{figure}

Let us return to Fig.1(c), and ask what is the correlation, if any, between
the vorticity of the current pattern and the direction of the magnetic
field. The answer is that the circulation around the central plaquette where
the vortex center reside is always opposite to that of the
magnetic field.\cite{han} (This is in contrast to the case in Fig.1(d).)
Moreover, we intentionally start with an
initial condition where the circulation of the central plaquette is the same
as that of the magnetic field, and find in the final
self-consistent solution that the vortex center moves by one lattice
spacing, so as to make the circulation of the central plaquette and the
magnetic field opposite. Thus the magnetic field pins the polarity of
the staggered current.

In Figs.2(a)-(d) we illustrate the evolution of the staggered current
pattern found at $10\%$ doping over a wide range of magnetic field. The
field strength is (a) $f=1/576$, (b) $f=1/144$, (c) $f=1/64$, and (d) $f=1/36
$, respectively. In physical units the fields considered here are very large
($f=1/1600$ roughly corresponds to a magnetic field of $10$ Tesla). It is
clear that as the vortices get closer their staggered current cores overlap
and permeate the entire lattice, as shown in Figs.2(c) and (d).\cite{leewen}
The end result is a state with uniform staggered current. This is a magnetic
field induced co-existing d-wave superconducting and staggered flux state.

Although the lowest field considered in this paper (roughly 30 Tesla) is
still very high from the experimental standpoint, none of our experimental
predictions (as will be discussed later)  will change qualitatively at a
lower field.

Just as in our earlier findings,\cite{han} the staggered current in the
mixed state also survives Gutzwiller projection.\cite{yokoyama} As an example, Fig.3 shows
the staggered current pattern at $x=6.25\%$ and $f=1/64$, before (a) and
after (b) the Gutzwiller projection. Although the staggered current is
weakened by the projection, it certainly does not disappear.

\begin{figure}
\epsfxsize=8cm\epsfysize=4cm\epsfbox{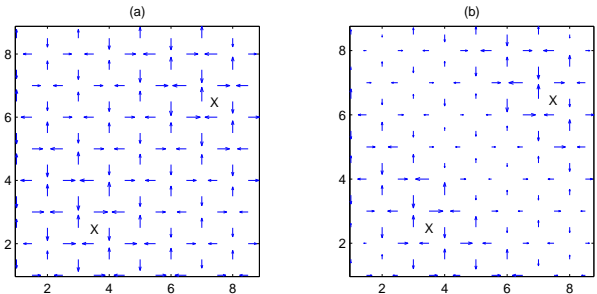}
\caption{Bond current pattern at $x=6.25\%$ and $f=1/64$. (a) Mean field
result. (b) Result after Gutzwiller projection. The crosses mark the
locations of the vortex cores.}
\end{figure}

\begin{figure}
\epsfxsize=8cm\epsfysize=7cm\epsfbox{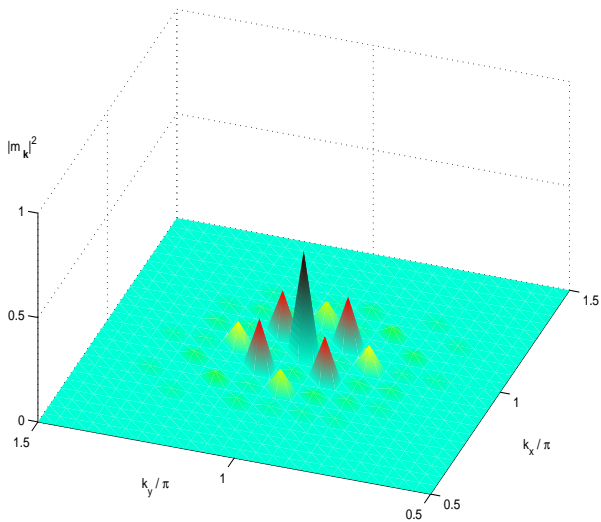}
\caption{$|m_{{\bf k}}|^{2}$ as a function of ${\bf k}$ corresponding to
Fig.2(a).}
\end{figure}

What are the experimental consequences of the staggered-current cores?
Figures 1(c) and
2 suggest that in a neutron scattering experiment there will be a magnetic
field induced Bragg peak near or even at $(\pi ,\pi )$. (Whether there is
a peak precisely at $(\pi ,\pi )$ depends on whether the vortex lattice is
entirely contained in one of the bi-partite sublattices of the underlying $%
CuO_{2}$ plane. The peaks around $(\pi ,\pi )$ should be more robust.) To
make sure that the scattering near $(\pi ,\pi )$ is induced by the vortex
core, one can 1) study the dependence of the intensity and position of the
Bragg peaks on the magnetic field and 2) correlate the peak position near $%
(\pi ,\pi )$ with that near $(0,0),$ which determines the structure of the vortex lattice.\cite{neutronstructure}
In Fig.4 we present $|m_{{\bf k}}|^{2}$ ,where $m_{{\bf k}}$ is the Fourier transform of the lattice curl of the
current ({\it i.e.}, the directed sum of the bond current around a plaquette). In this case the vortex lattice is
commensurate with one of the bi-partite sublattice of the underlying
crystal structure. We expect elastic neutron scattering will show
a similar pattern.

Seeing the magnetic field induced Bragg peaks near $(\pi,\pi)$ might be also consistent with vortex cores being
spin-antiferromagnetic.\cite{so5} Although it would be difficult to distinguish between the staggered orbital
moments and staggered spin moments from the standpoint of neutron scattering, the latter is much less likely from
theoretical considerations. Due to the presence of the external magnetic field and the vortex lattice,
all symmetries that ensure the degeneracy between the two polarity of the staggered flux order parameter are
broken. Thus the staggered flux in the vortex core is an induced, rather than spontaneous, order. This is not
true for the spin antiferromagnetic order. Indeed, under the experimental condition, rotation of spins around the
magnetic field direction remains an unbroken symmetry. As a result spin antiferromagnetic alignment in the
vortex core requires a spontaneous symmetry breaking, which is not possible for a finite (small) system such as
the core of a vortex.

\begin{figure}
\epsfxsize=8cm\epsfysize=7cm\epsfbox{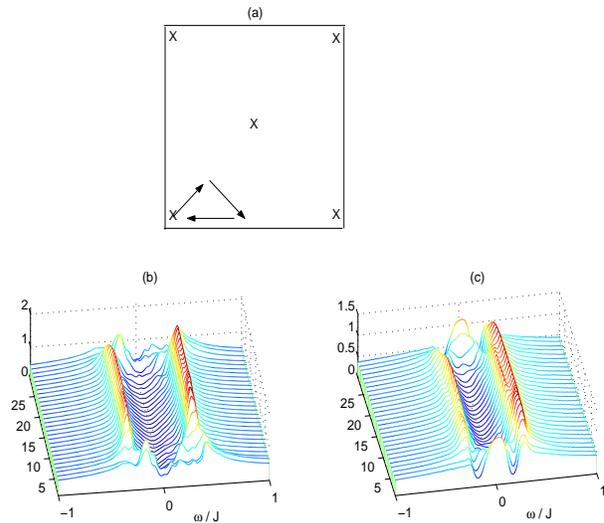}
\caption{(a) Schematic plot of the vortex lattice, and the the path along
which the density of states is calculated. The path begins with one of the
vortex cores (crosses). (b) The density of states with $x=10\%$ along the
path (starting from the front). (c) The same as (b) for $x=15\%$.}
\end{figure}

In addition to the scattering experiment staggered vortex core might also cause detectable difference in the local
tunneling spectra. In Fig.5 we present the local tunneling density of state as a function of location. The arrows
in (a) indicate the path (crosses mark the location of vortices) along which the local density of states is
computed. Figure 5(b) is the result for $x=10\%$ where there is staggered current in the vortex core. Figure 5(c)
is the result for $x=15\%$, where staggered current does not exist. The most significant difference occur at low
energies. While there is a zero-bias peak in the core of the $15\%$ vortex, there is no such peak in the core of
the $10\%$ vortex.\cite{stm} Since the staggered current is known to open up a pseudogap the result is not
surprising.

In general spin polarized tunneling experiment can be used to differentiate orbital versus spin
antiferromagnetism. In the presence of the latter, the spin-dependent tunneling density of state should show a
two-sublattice structure. We do not expect such effect for orbital antiferromagnetism,  because a) the Zeeman
splitting caused by the orbital moment is extremely small, and b) on a given site the staggered magnetic field
originated from four neighboring plaquettes tends to cancel.

If the signatures of the staggered current core discussed above are seen in
both the neutron and STM experiments it will constitute an extremely strong
evidence for the existence of staggered current in the vortex core.

Finally, what do we learn from the existence/non-existence of the staggered
current in the vortex core? From the beginning of high-$T_{c}$ physics, the
t-J model has been identified as the model that captures the competing local
interactions, {\it i.e.} charge hopping and spin antiferromagnetic exchange,
of the cuprate materials. During the last fifteen
years important progress has been
made on numerically simulating this model. Nonetheless questions such as
whether the ground state is homogeneous, and whether it is d-wave
superconducting still remain controversial. On the other hand the mean-field
theory of the t-J model has produced very tantalizing results. For example,
it predicts that d-wave pairing is among the most pronounced ordering
tendency of the model.\cite{kotliarliu} Recently, to a limited extent,
mean-field theory has also predicted the presence of stripe inhomogeneity.
\cite{stripe} In our opinion these successes give enough motivation
to check whether another robust prediction of the mean-field theory, the
close competition of the staggered flux phase with d-wave pairing, is
actually correct. The presence of staggered current in the core of a vortex
is a direct consequence of this competition. If such an ordered current
pattern is observed, it is reasonable to argue that we have understood the
rudimentary ordering tendency of the model. On the other hand, in the face
of negative experimental results we should seriously worry about the
validity of the mean-field theory, or perhaps even about our understanding
of the relevant local interactions.

\acknowledgments{QHW is supported by the National Natural Science
Foundation of China and the Ministry of Science and
Technology of China (NKBSF-G 19990646), and in part by the Berkeley
Scholars Program. DHL is supported by NSF grant DMR 99-71503.}

\widetext


\begin{references}
\bibitem{kotliar}  G. Kotliar, Phys. Rev. B {\bf 37}, 3664 (1988).

\bibitem{affleck}  I. Affleck, and J. B. Marston, Phys. Rev. B {\bf 37},
3774 (1988); J. B. Marston, and I. Affleck, Phys. Rev. B {\bf 39}, 11538
(1989); T. C. Hsu, J. B. Marston, and I. Affleck, Phys. Rev. B {\bf 43},
2866 (1991).

\bibitem{patrick} D. A. Ivanov, P. A. Lee, and Xiao-Gang Wen,
Phys. Rev. Lett. {\bf 84}, 3958 (2000).

\bibitem{clmn}  S. Chakravarty, R. B. Laughlin, D. K. Morr and C. Nayak,
cond-mat/0005443.

\bibitem{wang}  Qiang-Hua Wang, Jung Hoon Han, and Dung-Hai Lee,
cond-mat/0011398.

\bibitem{sudip}  Sumanta Tewari, Hae-Young Kee, Chetan Nayak, and Sudip
Chakravarty, cond-mat/0101027.

\bibitem{nayak}  Sudip Chakravarty, Hae-Young Kee, and Chetan Nayak,
cond-mat/0101204.

\bibitem{mook}  H. Mook, P. Dai and F. Dogan, submitted to Phys. Rev. Lett.

\bibitem{keimer}  Y. Sidis, {\it et al}, cond-mat/0101095.

\bibitem{leewen}  P. A. Lee, and Xiao-Gang Wen, cond-mat/0008419.

\bibitem{han}  Jung Hoon Han, Qiang-Hua Wang, and Dung-Hai Lee,
cond-mat/0012450.

\bibitem{macdonald}  Y. Wang, and A. H. MacDonald, Phys. Rev. B {\bf 52},
3876 (1995).

\bibitem{fluxquantum}  S. Sachdev, Phys. Rev. B {\bf 45}, 389 (1992); N.
Nagaosa, and P. A. Lee, Phys. Rev. B {\bf 45}, 966 (1992); M. Franz, and Z.
Tesanovic, cond-mat/0002137.

\bibitem{yokoyama} For the algorithm, see, {\it e.g.},
H. Yokoyama, and H. Shiba, J. Phys. Soc. Jpn. {\bf 56}, 3570 (1987).
Some straightward extension is needed to deal with the vortex lattice.

\bibitem{neutronstructure} M. Yethiraj, {\it et al}, Phys. Rev. Lett. {\bf 71},
3019 (1993); B. Keimer, {\it et al}, Phys. Rev. Lett. {\bf 73}, 3459 (1994).

\bibitem{so5} D. P. Arovas, A. J. Berlinsky, C. Kallin, and S.-C. Zhang,
Phys. Rev. Lett. {\bf 79}, 2871 (1997).

\bibitem{stm} I. Magio-Aprile, {\it et al}, Phys. Rev. Lett. {\bf 75},
2754 (1995); Ch. Renner, {\it et al}, Phys. Rev. Lett. {\bf 80},
3606 (1998); S. H. Pan, {\it et al}, Phys. Rev. Lett. {\bf 85}, 1536 (2000).

\bibitem{kotliarliu}  G. Kotliar, and J. Liu, Phys. Rev. B {\bf 38}, 5142
(1988).

\bibitem{stripe} Matthias Vojta, and S. Sachdev, Phys. Rev. Lett. {\bf %
83}, 3916 (1999); Jung Hoon Han, Qiang-Hua Wang, and Dung-Hai Lee,
cond-mat/0006046.
\end{references}
\end{document}